\documentclass[format=acmsmall, screen=true, review=false]{acmart}
\usepackage{booktabs} 
\usepackage{url}
\usepackage{latexsym}
\usepackage{amssymb}
\usepackage{algorithmic}
\usepackage[ruled,vlined]{algorithm2e}
\usepackage{tikz}
\usepackage[]{balance}
\usepackage{verbatim}
\usetikzlibrary{arrows,shapes}
\usepackage{multirow}
\usepackage{amsmath}
\usepackage{enumerate}
\usepackage[]{comment}
\usepackage{calrsfs}
\usepackage{mathtools}
\usepackage[]{subcaption}
\usepackage{float}
\DeclareMathOperator \real{\mathbb{R}}
\newcommand{\Ni}{({\em i})~}
\newcommand{\Nii}{({\em ii})~}

\DeclareMathAlphabet{\pazocal}{OMS}{zplm}{m}{n}

\newcommand{\Ts}{\pazocal{T}}
\newcommand{\Gs}{\pazocal{G}}

\usepackage{booktabs} 
\setcopyright{rightsretained}

\newcommand{\ra}[1]{\renewcommand{\arraystretch}{#1}}

\acmDOI{10.475/123_4}

\acmISBN{123-4567-24-567/08/06}

\acmConference[KDD]{ACM }{August, 2018}{London, United Kingdom} 
\acmYear{2017}
\copyrightyear{2016}

\acmPrice{15.00}

\begin{document}
\title [Temporal Smoothness] {Models for Capturing Temporal Smoothness in Evolving Networks for Learning Latent Representation of Nodes}

\author[Saha, Williams, Hasan]{Tanay Kumar Saha, Thomas Williams and Mohammad Al Hasan}
\affiliation{%
   \institution{Indiana University Purdue University}
   \streetaddress{750 W. Michigan St}
   \city{Indianapolis} 
   \state{IN} 
   \postcode{46202}
 }
\email{tksaha, alhasan, tfwillia@iupui.edu}

\author[Joty]{Shafiq Joty}
\affiliation{%
  \institution{Nanyang Technological University}
  \streetaddress{50 Nanyang Ave}
  \state{Singapore} 
  \postcode{639798}
}
\email{srjoty@ntu.edu.sg}
\author[Varberg]{Nicholas K. Varberg}
\affiliation{%
   \institution{Wheaton College}
   \streetaddress{501 College Ave}
   \city{Wheaton} 
   \state{IL} 
   \postcode{60187}
 }
\email{nick.varberg@my.wheaton.edu}

 
\begin{abstract}
In a dynamic network, the neighborhood of the vertices evolve across
different temporal snapshots of the network. Accurate modeling of this temporal evolution 
can help solve complex tasks involving real-life social and interaction networks. 
However, existing models for learning latent representation  are inadequate
for obtaining the representation vectors of the vertices for different time-stamps of
a dynamic network in a meaningful way. In this paper, we propose latent representation
learning models for dynamic networks which overcome the above limitation by considering two different kinds of temporal smoothness: \Ni retrofitted, 
and \Nii linear transformation. The retrofitted model tracks the representation vector
of a vertex over time, facilitating vertex-based temporal analysis of a network. On the other hand, 
linear transformation based model provides a smooth transition operator which maps 
the representation vectors of all vertices from one temporal snapshot to the next (unobserved)
snapshot---this facilitates prediction of the state of a network in a future time-stamp. We validate 
the performance of our proposed models by employing them for solving the temporal link prediction task.
Experiments on 9 real-life networks from various domains validate that the proposed models are significantly better than the existing models for predicting the dynamics 
of an evolving network.

\end{abstract}



\maketitle

\section{Introduction}
\label{sec:tempn2v:intro}

Accurate modeling of temporal evolution of networks can help solve 
complex tasks involving social and interaction networks. For instance, capturing temporal dynamics of user interactions can explain how communities are formed and dissolved in a network over time. Temporal 
co-movement of financial asset prices explains how financial assets are clustered pronouncedly during an economic downturn, causing a cascading effect that leads to financial crisis. Temporal network models
can explain the way social network topologies facilitate (or inhibit) grievances to intensify
collective organization, leading to imminent crisis and conflict in a community~\cite{Korolov.Lu.ea:2016}. Unfortunately, most of the
network based analyses consider only the network topology~\cite{Perozzi.Al-Rfou.ea:14,Grover.Leskovec:16,Tang.Qu.ea:15*2}, a few consider nodal attributes and 
topology~\cite{Huang.Li.ea:2017}, and almost all ignore the temporal evolution of a network. The objective of this work is to capture temporal evolution of networks by learning latent representation of vertices over time.

Over the past few years, there has been a surge in research~\cite{Perozzi.Al-Rfou.ea:14,Grover.Leskovec:16,Tang.Qu.ea:15*2,Wang.Cui:16,Wang.Cui:17} on embedding the vertices of a network into a low-dimensional, dense vector space. These embedding models utilize the topological information of a network to maximize objective functions that capture the notion that nodes with similar topological arrangements should be distributed closely in the learned low-dimensional vector space. The embedded vector representation of the vertices in such a vector space enables effortless invocation of off-the-shelf machine learning algorithms, thereby facilitating several downstream network mining tasks, including node classification~\cite{Tu.Zhang.ea:16}, link prediction~\cite{Grover.Leskovec:16}, and community detection~\cite{Wang.Cui:17}. However, most of the existing network embedding methods, including DeepWalk~\cite{Perozzi.Al-Rfou.ea:14}, LINE~\cite{Tang.Qu.ea:15*2}, and Node2Vec~\cite{Grover.Leskovec:16} only consider a static network in which the time-stamp of the edges are ignored. 
The embedding vectors of the nodes do not have any temporal connotation. 
These time-agnostic models may produce incorrect analysis---for example, in a static link-prediction task, node vectors might have been learned (inadvertently) by using future edges, but foreseeing future edges is impossible in a real-time setup.
In summary, temporal network models should learn latent representation of vertices by considering the edges in their temporal order to make the model interpretable along the time axis, leading to the discovery of temporal evolution patterns of a dynamic network.

If a temporal network is represented as a collection of snapshots at discrete time intervals, one may attempt to
use static network models (e.g., LINE, Node2Vec) for learning vector representation of vertices at each time-stamp independently. Through the embedding vectors of each vertices, these models encode useful semantic information, specifically, proximity and homophily relation among the vertices. But, the learning is limited to only one given time-stamp. More importantly,
due to the independence in learning process across the time-stamps, the latent vectors of the vertices are
embedded in different affine spaces for different temporal snapshots of the network. Therefore, there is no temporal mapping across the affine spaces to connect the embedding vectors of the same vertex across different time-stamps. 
Another related objective that we may have is to capture the temporal progression of the vertices in a latent space, say, for solving the task of community evolution over time; existing embedding models for static networks also fail to fulfill this objective. 

We want to emphasize the differences between the two objectives which we have discussed in the above paragraph. For the first objective, we want an operator to transform the coordinates of the identical vertices (as obtained from the embedding of a dynamic graph in different time-stamps) from one affine space to another affine space. 
We refer to this objective as {\em global temporal smoothness} as we achieve this by considering all the vertices of a network holistically. 
Transformation here acts as a smoothness operator to connect the embedding vectors of the vertices over the time space.
On the other hand, for the second objective, we apply temporal smoothness over the vertices independently to ensure that their vectors have a smooth progression through the time-stamps.
We call this {\em local temporal smoothness}. 
The existing network models fail to return temporal representation vectors of the vertices fulfilling either of the objectives; overcoming this limitation is the main motivation of this work. Note that,
some earlier works have used temporal smoothness for evolutionary clustering~\cite{Chi.Song.ea:2007} and link prediction in a dynamic network~\cite{Zhu.Guo:16}. Both works use an identical objective function which minimizes a matrix-factorization coupled with a temporal smoothing regularization. But, these models only use first-order proximity. To the best of our knowledge, no models consider higher-order proximities and temporal smoothness to provide latent representation of vertices for each time-stamp of a dynamic network.

\begin{figure}
\centering
\begin{subfigure}[]{0.4\textwidth}
  \includegraphics[width=3in]{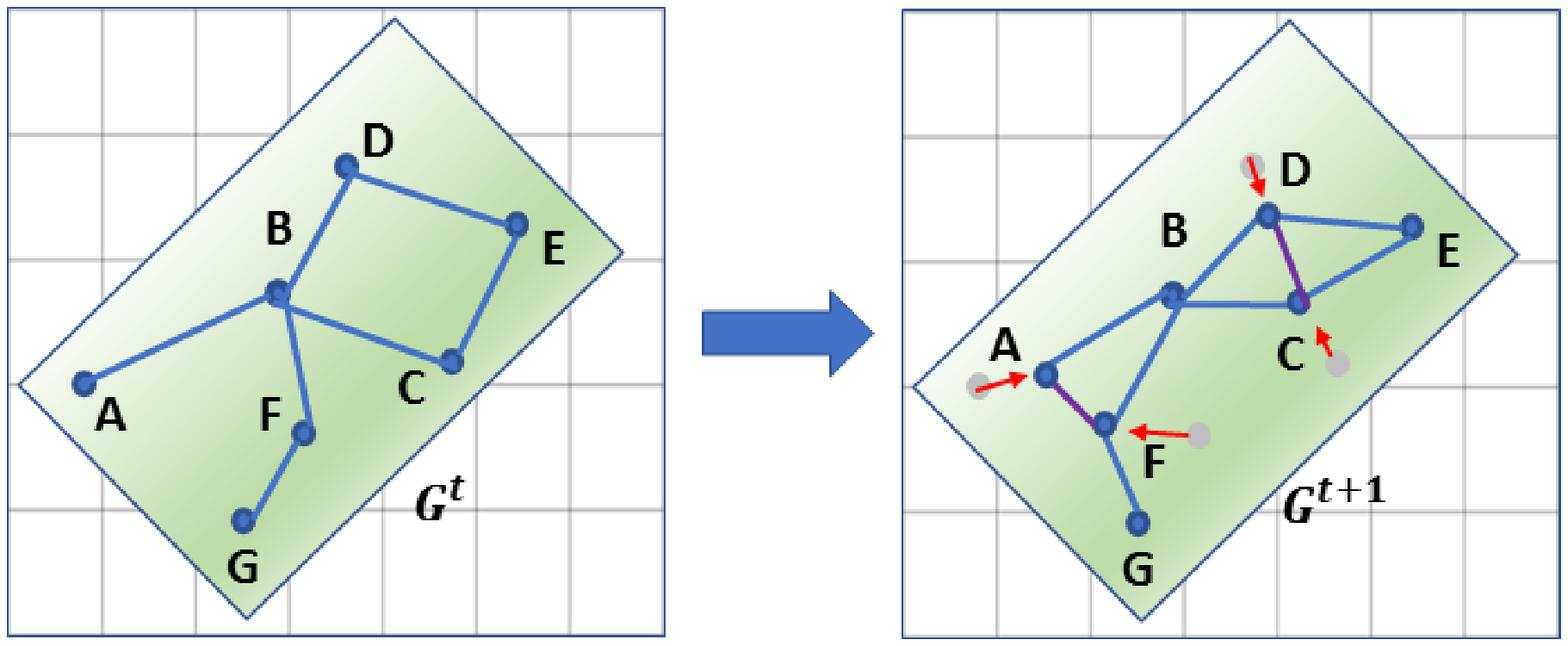}
\end{subfigure}
\hrule
\begin{subfigure}[]{0.4\textwidth}
  \includegraphics[width=3in]{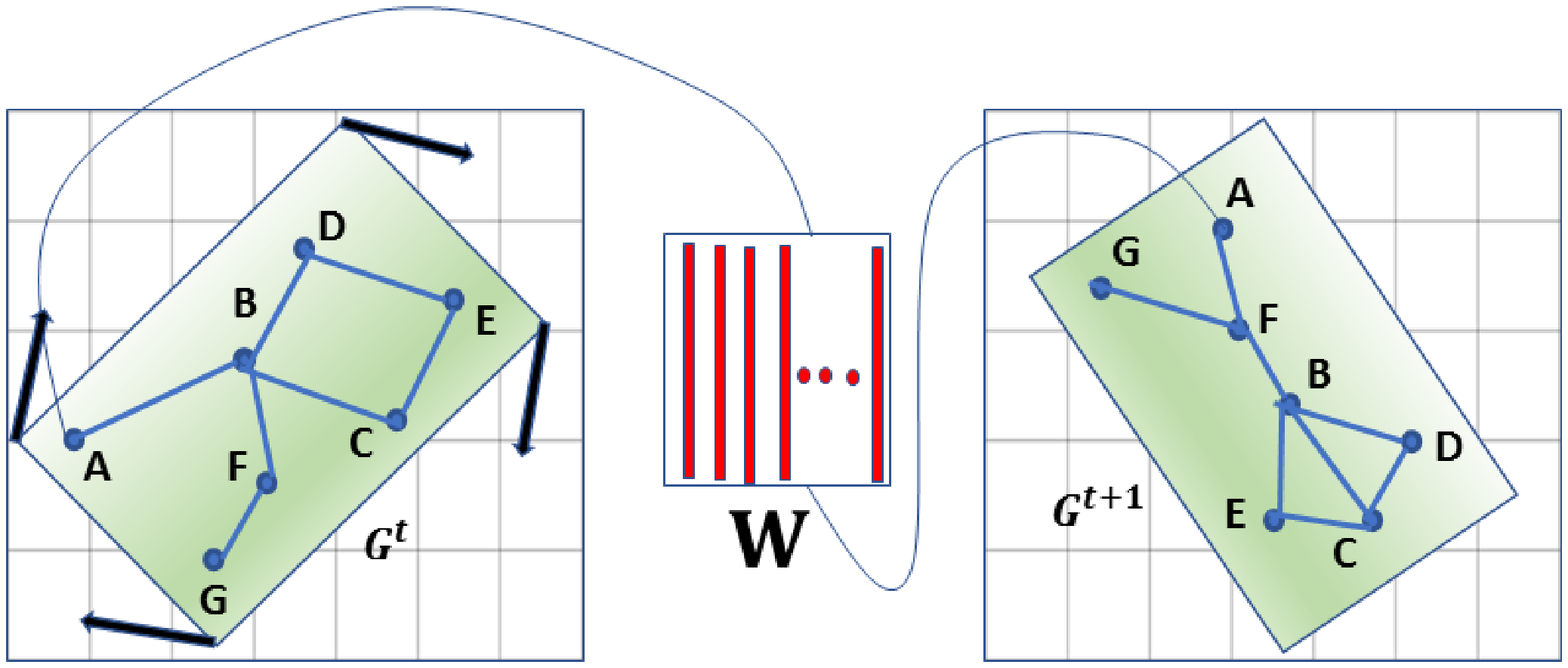}
\end{subfigure}
\caption{A conceptual sketch of retrofitting (top) and linear transformation (bottom) based temporal smoothness.}
\label{fig:concept}
\end{figure}

In this work, we propose two embedding models, \Ni retrofitted, and \Nii linear transformation, each fulfilling one of the smoothness objectives. Figure~\ref{fig:concept} gives a conceptual demonstration of these models. The \emph{retrofitted} model satisfies the local temporal smoothness objective  by assuming that the evolution of the network is vertex-centric. In each time stamp, a small fraction of the vertices experience changes in their neighborhood. The {\em retrofitted} model  
smoothly updates (retrofits) the embedding vectors of vertices, which are attached to the new edges in a given time stamp. As shown in the top example of Figure~\ref{fig:concept} the presence of new edges $AF$ and $CD$ in the graph $G^{t+1}$, updates the vector representation of the vertices $A, C, D$ and $F$ from their prior position corresponding to $G^{t}$. For the first time-stamp though, the model employs an existing latent representation model (neural network, manifold learning, or matrix-factorization based like PCA or SVD) to learn the representation vector of the vertices, but for subsequent
time-stamps the position of the vectors are updated by a local update method as discussed above.
The retrofitted model enables temporal tracking of the vertices of a network which can be instrumental for solving an evolutionary clustering of the vertices or to discover the evolution of communities over time. 

Our second model, the \emph{linear transformation} model assumes that the network evolution over time is a global process, which
makes the evolution network-centric, instead of vertex-centric. To accommodate this assumption, this model attempts to fulfill
the global temporal smoothness objective by considering the temporal evolution of a network as a linear transformation operator over the vertex embedding vectors of successive time-stamps, as shown in Figure~\ref{fig:concept}(bottom). In this figure
we show the (independently learned) embedding vectors of the vertices in $G_t$ and $G_{t+1}$, and our objective
is to learn the transformation operator $\mathbf{W}$, which can map the vectors
of each vertices from its position in $G_t$ to its position in $G_{t+1}$.
Once learned, the operator $\mathbf{W}$ 
is able to map the latent representation from a known snapshot to the next (unobserved) snapshot. The model first obtains embedding functions of all temporal snapshots of the graph and then it learns
the transformation operation which best explains the evolution of vertex embedding vectors across different time-stamps. 
We explore two ways to learn the transformation matrix: homogeneous and heterogeneous. In \emph{homogeneous mapping}, we assume that the transformation operation is  the same across any two successive time-stamps. 
So, we learn a single (shared) transformation matrix that maps the representation from a snapshot to the next snapshot. \emph{Heterogeneous mapping} on the other hand refrains from the uniformity assumption, 
considering that every pair of time-stamps has a different transformation geometry. So, the model learns a projection matrix for every subsequent time-pairs and then combines them while performing smoothing in the time dimension. 

Contributions of this paper are summarized as below.
\begin{enumerate}
\item We propose two novel models for learning the vertex representation of a dynamic network having many temporal snapshots. In these models we introduce two different
kinds of temporal smoothness concepts: global and local, which complement each other.

\item We validate our proposed models by utilizing them for solving the temporal link prediction task on nine different datasets from three different domains: citation, social, and messaging. Experimental results show that when compared against an
existing state-of-the-art temporal smoothness based dynamic link prediction model, for all datasets our proposed
methods improve the link prediction performance by values ranging from 0.20 to 0.60 on different metrics, such as, AUC, PRAUC, and NDCG.

\item We made our code, datasets, and experimental setups publicly available\footnote{https://gitlab.com/tksaha/temporalnode2vec.git} to support the spirit of reproducible research and to enable further development in this area.

\end{enumerate}

\section{Related Work}
\label{sec:tempn2v:relwork}
Recently, models for learning latent representation of vertices \cite{Perozzi.Al-Rfou.ea:14,Tang.Qu.ea:15*2,Wang.Cui:16,
Grover.Leskovec:16, Benson.Gleich:16, kipf.welling:17, Tu.Zhang.ea:16} of a static networks have become very popular. 
These models vary in the way they exploit information from the static network as they learn the low-dimensional representation of the vertices.
For example, DeepWalk \cite{Perozzi.Al-Rfou.ea:14} and Node2Vec \cite{Grover.Leskovec:16} design random walks to find the node pairs that should be considered similar, and then use word2vec's skip-gram model \cite{Mikolov.Le.Sutskever:13} to learn representations. LINE \cite{Tang.Qu.ea:15*2} extracts two kinds of proximities among the nodes: \Ni direct link (first-order), and \Nii structural (second-order) proximity.
SDNE \cite{Wang.Cui:16} also captures first and second-order proximity, but it uses an encoder-decoder framework and Laplacian regularization to capture the proximity between a pair of vertices. Benson's model \cite{Benson.Gleich:16} extracts proximity among the nodes using network motifs. Qiu et al.\cite{netmf} show that DeepWalk, LINE, and Node2Vec models can be unified under the matrix factorization framework.
Manifold learning models, such as ISOMAP and Locally Linear
Embedding (LLE), also aim to reduce dimensionality.
ISOMAP \cite{tenenbaum2000global} uses the geodesic distances among the nodes to learn a low-dimensional vector representation for each node. LLE \cite{roweis2000nonlinear} eliminates ISOMAP's need to estimate the pairwise 
distances between widely separated nodes. The model assumes that each node and its 
neighbors lie on or near a locally linear patch of a manifold and subsequently, learns a neighborhood preserving 
latent space representation by locally linear reconstruction.



Although we found many embedding methods for static networks, we found only one related work for dynamic networks. Zhu et al. \cite{Zhu.Guo:16} attempt dynamic link-prediction by adding a temporal-smoothing regularization term to a non-negative matrix factorization objective. Their goal is to reconstruct the adjacency matrix of different time-stamps of a graph. They use a Block-Coordinate Gradient Descent (BCGD) algorithm to perform non-negative factorization. Their
formulation is almost identical to Chi et al. \cite{Chi.Song.ea:2007}
who perform evolutionary spectral clustering that captures temporal smoothness. Because matrix factorization provides embedding vectors
of the nodes for each time-stamp, the factorization by-product from this work can be 
considered as dynamic network embeddings. In experiment section we compare our proposed methods with this work.

 
There are few other works which model dynamic networks or solve link prediction on dynamic networks. But, 
they do not learn latent vectors of the vertices for each time-stamps.
For instance, the method in \cite{Gunes2015} computes a number of different node similarity scores by summing
those similarities with weights learned for different time-stamps of the network.  
Rahman et al.~\cite{Rahman.Hasan:16} use graphlet transitions over two successive
snapshots to solve the dynamic link prediction problem.
A deep learning solution is proposed in \cite{Li:2014}, which uses a 
collection of Restricted Boltzmann Machines with neighbor 
influence for link prediction in dynamic networks. 
Matrix and tensor factorization based solutions are presented in \cite{Dunlavy:2011,KevinS:2015}.
\section{Problem Formulation} \label{sec:prob}

Let $\Ts = \{1, 2, \ldots, T\}$ be a finite set of time-stamps for an evolving (undirected) network $G$, and for
$t \in [1,T]$, $G_t = (V_t, E_t)$ denotes the network state at time $t$ with $V_t$ being the set of vertices and $E_t$ being the set of edges of graph $G_t$ at $t$'th time-stamp. The sequence of network snapshots is thus represented by $\Gs$ = ($G_1$, $G_2$,  \ldots, $G_T$).  
For simplicity, we assume that all the networks in $\Gs$ have the same vertex set, i.e., $G_t$ = $(V, E_t)$ for $t=1, 2, \ldots, T$.\footnote{This assumption is not a limitation, as the proposed models can easily be adapted for the
case when this assumption does not hold.} We also assume that apart from the link information in a network, no other attribute data for the nodes or edges are available.

Now, let $\phi_{t}:V \rightarrow \real^{d}$ be the mapping function at time-stamp $t$ that returns the distributed representations, i.e., real-valued $d$-dimensional vectors representing the vertices in $G_t$. In terms of data structure, $\phi_t$ is simply a look-up matrix of size $|V| \times d$, where $|V|$ is the total number of vertices in the network. The task of dynamic network embedding is to approximate $\phi_t$ from the sequence of first $t$ network snapshots, represented as, $\Gs_t$ = ($G_1$, $G_2$, \ldots, $G_{t})$.
Unlike existing embedding models, for learning embedding function $\phi_t$, we want to utilize both the topological information in $G_{t}$ and the trends in temporal dynamics exhibited by the sequence of network snapshots up to time $t$.

In this work, we propose two different models: retrofitted, and linear projection, each feeding on a specific temporal smoothness assumption.

\begin{enumerate}

\item For the retrofitted model, we first learn $\phi_{1}$ from the network information in $G_{1}$ using any of the state-of-art static network embedding methods (e.g., Node2Vec, DeepWalk). Then we capture the temporal network dynamics by retrofitting $\phi_{1}$ successively with the network snapshots $G_2$, $G_3$, \dots, $G_{t}$. 


\item For the linear transformation model, we learn a linear transformation matrix $W \in \real^{d \times d}$ to map $\phi_{t-1}$ to $\phi_{t}$. The matrix $W$ is trained after all the $\phi_{i}$ for $1 \le i \le t$ are obtained by using
one of the existing static embedding methods on network snapshots in $\Gs_t = (G_1, G_2, \ldots, G_{t})$.


\end{enumerate}

To validate our proposed models we use temporal link prediction as an example task,
where we predict the links in a future snapshot of a network, namely $G_{T+1}$. However, we would like to point out that our proposed embedding methods are agnostic to the task at hand. \\




\section{Method}
\label{sec:tempn2v:Method}

In the following, we describe our proposed models in detail. 

\subsection{Retrofitted Model} \label{sec:retro_node2vec}
\usetikzlibrary{calc}
\usetikzlibrary{decorations.pathreplacing,angles,quotes}
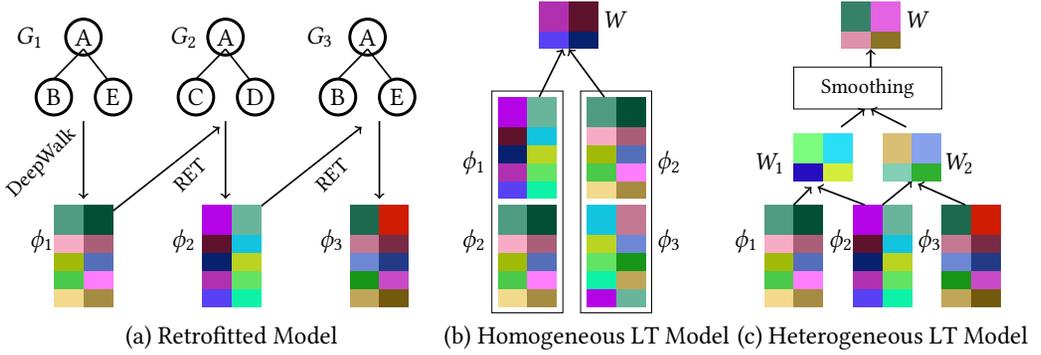
\begin{figure}
\pgfmathsetseed{1}
\resizebox{\textwidth}{!}{
\begin{tikzpicture}
\definecolorseries{test}{rgb}{step}[rgb]{.95,.85,.55}{.17,.47,.37}
\def\startx{12}
\def\starty{0}

\resetcolorseries[35]{test}
\foreach \x in {\startx,\startx+0.5}{
    \foreach \y in {\starty,\starty+0.3,\starty+0.6,\starty+0.9,\starty+1.2}
        \fill[test!!+] (\x,\y) rectangle ++(0.5,0.5);
    } 
\foreach \x in {\startx+1.5,\startx+2.0}{
    \foreach \y in {\starty,\starty+0.3, \starty+0.6, \starty+0.9, \starty+1.2}
        \fill[test!!+] (\x,\y) rectangle ++(0.5,0.5);
    }
\foreach \x in {\startx+3.0, \startx+3.5}{
    \foreach \y in {\starty,\starty+0.3, \starty+0.6, \starty+0.9, \starty+1.2}
        \fill[test!!+] (\x,\y) rectangle ++(0.5,0.5);
    }
\draw[thick,->] (\startx+0.5,\starty+1.7) -- (\startx+0.8,\starty+2.0);
\draw[thick,->] (\startx+1.7,\starty+1.7) -- (\startx+0.9,\starty+2.0);
\draw[thick,->] (\startx+2.0,\starty+1.7) -- (\startx+2.5,\starty+2.1);
\draw[thick,->] (\startx+3.4,\starty+1.7) -- (\startx+2.6,\starty+2.05);



\node (d) at (\startx -0.3,\starty+1.1) {\Large $\phi_1$};  
\node (d) at (\startx+1.3,\starty+1.1) {\Large $\phi_2$};  
\node (d) at (\startx+2.8,\starty+1.1) {\Large $\phi_3$}; 

\node (d) at (\startx+0.1,\starty+2.4) {\Large $W_1$}; 
\node (e) at (\startx+3.3,\starty+2.4) {\Large $W_2$}; 

\node (f) at (\startx+2.6,\starty+4.8) {\Large $W$}; 
 
\foreach \x in {\startx+0.5, \startx+1.0}{
    \foreach \y in {\starty+2.1, \starty+2.4}
        \fill[test!!+] (\x,\y) rectangle ++(0.5,0.5);
    }   
    
\foreach \x in {\startx+2.0, \startx+2.5}{
    \foreach \y in {\starty+2.1, \starty+2.4}
        \fill[test!!+] (\x,\y) rectangle ++(0.5,0.5);
    }   
\draw (\startx+0.5,\starty+3.3) rectangle (\startx+3.0,\starty+4.0) node[pos=.5] {Smoothing};

\draw[thick,->] (\startx+1.3,\starty+3.0) -- (\startx+1.7,\starty+3.3);
\draw[thick,->] (\startx+2.4,\starty+3.0) -- (\startx+1.8,\starty+3.3);

\draw[thick,->] (\startx+1.8,\starty+4.0) -- (\startx+1.8,\starty+4.3);

\foreach \x in {\startx+1.3, \startx+1.8}{
    \foreach \y in {\starty+4.3, \starty+4.6}
        \fill[test!!+] (\x,\y) rectangle ++(0.5,0.5);
    }  
    
\node (c) at (14.0, -0.2in) { \Large (c) Heterogeneous LT Model};    

    
\def\startx{3}
\def\starty{0}


\node () at (\startx+4.1,\starty+1.1) {\Large $\phi_2$};  
\node () at (\startx+4.1,\starty+2.4) {\Large $\phi_1$}; 
\node () at (\startx+7.4,\starty+1.1) {\Large $\phi_3$};  
\node () at (\startx+7.4,\starty+2.4) {\Large $\phi_2$}; 
\node (f) at (\startx+6.5,\starty+4.8) {\Large $W$};

\draw (7.4,-0.1) rectangle (8.6,3.6);
\draw (8.9,-0.1) rectangle (10.1,3.6);

\resetcolorseries[35]{test}   
\foreach \x in {\startx+4.5, \startx+5.0}{
   \foreach \y in {\starty,\starty+0.3,\starty+0.6, \starty+0.9, \starty+1.2}
        \fill[test!!+] (\x,\y) rectangle ++(0.5,0.5);
    }
   \foreach \x in {\startx+4.5, \startx+5.0}{
    \foreach \y in {\starty+1.8, \starty+2.1, \starty+2.4, \starty+2.7, \starty+3.0}
        \fill[test!!+] (\x,\y) rectangle ++(0.5,0.5);
    }
   
\resetcolorseries[35]{test}
    \foreach \x in {\startx+6.0, \startx+6.5}{
      \foreach \y in {\starty+1.8, \starty+2.1,\starty+2.4, \starty+2.7, \starty+3.0}
        \fill[test!!+] (\x,\y) rectangle ++(0.5,0.5);
     }
\foreach \x in {\startx+5.2, \startx+5.7}{
    \foreach \y in {\starty+4.3, \starty+4.6}
        \fill[test!!+] (\x,\y) rectangle ++(0.5,0.5);
    }   
\foreach \x in {\startx+6.0, \startx+6.5}{
   \foreach \y in {\starty, \starty+0.3, \starty+0.6, \starty+0.9, \starty+1.2}
        \fill[test!!+] (\x,\y) rectangle ++(0.5,0.5);
    }
    
\draw[thick,->] (\startx+5.2,\starty+3.5) -- (\startx+5.6,\starty+4.3);
\draw[thick,->] (\startx+6.3,\starty+3.5) -- (\startx+5.7,\starty+4.3);

\def\startx{0}
\def\starty{0}
\resetcolorseries[55]{test} 
\newcommand*{\XAxisMin}{0}%
\newcommand*{\XAxisMax}{6}%
\newcommand*{\OriginX}{0}%
\newcommand*{\OriginY}{0.0}%
\newcommand*{\IncrY}{1.0}
\newcommand*{\IncrX}{0.1}
\newcommand*{\Xextension}{6}
\newcommand*{\Yextension}{1}
\newcommand*{\IncrYDM}{0.7}

\def\startxx{0}
\def\startyy{0}

\node () at (\startx - 0.2,\starty+1.1) {\Large $\phi_1$};  
\node () at (\startx + 2.2,\starty+1.1) {\Large $\phi_2$}; 
\node () at (\startx + 4.7,\starty+1.1) {\Large $\phi_3$}; 

\node () at (\startx - 0.4,\starty+4.5) {\Large $G_1$};  
\node () at (\startx + 2.2,\starty+4.5) {\Large $G_2$};
\node () at (\startx + 4.5,\starty+4.5) {\Large $G_3$};

\draw [black,thick, line width=0.5mm] (\OriginX+\startxx+0.5,\OriginY+\startyy+4.5) circle (0.3cm) node[] at (\OriginX+\startxx+0.5,\OriginY+\startyy+4.5) {\Large A};

\draw [black,thick, line width=0.5mm] (\OriginX+\startxx+0.0,\OriginY+\startyy+3.5) circle (0.3cm) node[] at (\OriginX+\startxx+0.0,\OriginY+\startyy+3.5) {\Large B};

\draw [black,thick, line width=0.5mm] (\OriginX+\startxx+1.0,\OriginY+\startyy+3.5) circle (0.3cm) node[] at (\OriginX+\startxx+1.0,\OriginY+\startyy+3.5) {\Large E};

\draw[thick,-] (\OriginX+\startxx+0.5,\OriginY+\startyy+4.3) -- (\startxx+0.9,\startyy+3.8);
\draw[thick,-] (\OriginX+\startxx+0.5,\OriginY+\startyy+4.3) -- (\startxx+0.0,\startyy+3.8);

\draw[thick,->] (\OriginX+\startxx+0.5,\OriginY+\startyy+3.1) -- (\startxx+0.5,\startyy+1.8);
\draw[thick,->] (\OriginX+\startxx+1.0,\OriginY+\startyy+1.6) -- (\startxx+2.8,\startyy+3.0);


\node (a) at (\OriginX+3.0, \OriginY-0.2in) { \Large (a) Retrofitted Model};
\node[rotate=45] at (\startxx-0.2,\startyy+2.4) {DeepWalk};
\node[rotate=45] at (\startxx+2.3,\startyy+2.2) {RET};

\def\startxx{2.4}
\def\startyy{0}

\draw [black,thick, line width=0.5mm] (\OriginX+\startxx+0.5,\OriginY+\startyy+4.5) circle (0.3cm) node[] at (\OriginX+\startxx+0.5,\OriginY+\startyy+4.5) {\Large A};

\draw [black,thick, line width=0.5mm] (\OriginX+\startxx+0.0,\OriginY+\startyy+3.5) circle (0.3cm) node[] at (\OriginX+\startxx+0.0,\OriginY+\startyy+3.5) {\Large C};

\draw [black,thick, line width=0.5mm] (\OriginX+\startxx+1.0,\OriginY+\startyy+3.5) circle (0.3cm) node[] at (\OriginX+\startxx+1.0,\OriginY+\startyy+3.5) {\Large D};

\draw[thick,-] (\OriginX+\startxx+0.5,\OriginY+\startyy+4.3) -- (\startxx+0.9,\startyy+3.8);
\draw[thick,-] (\OriginX+\startxx+0.5,\OriginY+\startyy+4.3) -- (\startxx+0.0,\startyy+3.8);
\draw[thick,->] (\OriginX+\startxx+0.5,\OriginY+\startyy+3.1) -- (\startxx+0.5,\startyy+1.8);

\draw[thick,->] (\OriginX+\startxx+1.0,\OriginY+\startyy+1.6) -- (\startxx+2.8,\startyy+3.0);

\node[rotate=45] at (\startxx+2.3,\startyy+2.2) {RET};

\def\startxx{4.8}
\def\startyy{0}
\draw [black, thick, line width=0.5mm] (\OriginX+\startxx+0.5,\OriginY+\startyy+4.5) circle (0.3cm) node[] at (\OriginX+\startxx+0.5,\OriginY+\startyy+4.5) {\Large A};

\draw [black,thick, line width=0.5mm] (\OriginX+\startxx+0.0,\OriginY+\startyy+3.5) circle (0.3cm) node[] at (\OriginX+\startxx+0.0,\OriginY+\startyy+3.5) {\Large B};

\draw [black,thick, line width=0.5mm] (\OriginX+\startxx+1.0,\OriginY+\startyy+3.5) circle (0.3cm) node[] at (\OriginX+\startxx+1.0,\OriginY+\startyy+3.5) {\Large E};

\draw[thick,-] (\OriginX+\startxx+0.5,\OriginY+\startyy+4.3) -- (\startxx+0.9,\startyy+3.8);
\draw[thick,-] (\OriginX+\startxx+0.5,\OriginY+\startyy+4.3) -- (\startxx+0.0,\startyy+3.8);
\draw[thick,->] (\OriginX+\startxx+0.7,\OriginY+\startyy+3.1) -- (\startxx+0.7,\startyy+1.8);

\foreach \x in {\startx,\startx+0.5}{
    \foreach \y in {\starty,\starty+0.3,\starty+0.6,\starty+0.9,\starty+1.2}
        \fill[test!!+] (\x,\y) rectangle ++(0.5,0.5);
    }
\foreach \x in {\startx+2.5,\startx+3.0}{
    \foreach \y in {\starty,\starty+0.3, \starty+0.6, \starty+0.9, \starty+1.2}
        \fill[test!!+] (\x,\y) rectangle ++(0.5,0.5);
       }
\foreach \x in {\startx+5.0,\startx+5.5}{
    \foreach \y in {\starty,\starty+0.3, \starty+0.6, \starty+0.9, \starty+1.2}
        \fill[test!!+] (\x,\y) rectangle ++(0.5,0.5);
       }

\node (a) at (\OriginX+9.0, \OriginY-0.2in) { \Large (b) Homogeneous LT Model};    
\end{tikzpicture}}
\caption{Toy illustration of our method. $\phi$'s represent the embedding vectors of the vertices. (a) for retrofitted model, we first learn $\phi_1$ by using any  of static embedding learning models. We then use retrofitting to learn $\phi_2$ and $\phi_3$ using ($\phi_1$, $G_2$) and ($\phi_2$, $G_3$), successively. (b) \& (c) for linear transformation (LT) models, we first learn $\phi_1$, $\phi_2$, and $\phi_3$ by using any of static embedding models, and use these embeddings to learn a transformation matrix $W$.  Please see Section \ref{sec:tempn2v:Method} for details about how to learn $W$ for homogeneous and heterogeneous transformation models.}
\label{fig:models}
\end{figure}



This model is based on the local temporal smoothness assumption, where the smoothness is applied to different vertices independently. This assumption is needed to track the embedding vector of the vertices as the network evolves. As shown in Figure \ref{fig:models}(a), we do not learn the mapping function, $\phi$, from different temporal snapshots of the graphs. Instead, we learn $\phi_1$ from $G_1$ by using any of the static network embedding models discussed in Section \ref{sec:tempn2v:relwork}. 
Then, we transform $\phi_{1}$ to $\phi_{t}$, by iteratively retrofitting information from later network snapshots $G_2, G_3, \ldots, G_{t}$. 

In retrofitting, we revise $\phi_{t-1} (v)$ by using the neighborhood information available from the graph snapshot at time $t$, so that the resulting vector  $\phi_t ({v})$ is similar to the prior vector $\phi_{(t-1)} ({v})$ and at the same time close to the vectors of its adjacent nodes in $G_t$. The similarity between $\phi_t(v)$ and $\phi_{t-1}(v)$ enables the vertex $v$ to move smoothly in the embedded space as time progresses from $t-1$ to $t$. The closeness of $\phi_t(v)$ with its neighbor at
time $t$ satisfies the proximity requirement of any static network embedding model. Thus, we minimize the objective function:


\begin{equation}
J(\phi_t) = \underbrace{\sum_{{v} \in V} \alpha_v ||\phi_{t} ({v}) - \phi_{(t-1)} ({v})||^2}_\text{Temporal Smoothing} + \hspace{-2mm} \underbrace{\sum_{({v},{u})\in E_t} \hspace{-1mm} \beta_{u,v} ||\phi_{t}({u}) - \phi_ {t}({v})||^2}_\text{Network Proximity}, \label{eq:ret}
\end{equation} 


\noindent where $\alpha$ controls the strength to which the algorithm matches the prior vectors, for supporting \emph{temporal smoothness}, and $\beta$ controls the emphasis on \emph{network proximity}. The quadratic cost in Equation \ref{eq:ret} is convex in $\phi_{t}$, and has a closed form solution \cite{Talukdar.Koby.09}. The closed form expression requires an inversion operation, which can be expensive for large networks. 
The Jacobi method, an online algorithm, is more efficient as it solves the problem iteratively. The Jacobi method utilizes the following update rule:

\begin{equation}
\phi_{t}({v}) \leftarrow  \frac{\alpha_{v} \phi_{(t-1)}({v}) +  \sum_{{u}} \beta_{{v}, {u}}  \phi_{t}({u})}{\alpha_{v} + \sum_{{u}} \beta_{{v}, {u}}} \label{eq:ret2}. 
\end{equation}

\begin{algorithm}[t]
\SetAlgoNoLine
\SetNlSkip{0em}
\SetKwInOut{Input}{Input}
\SetKwInOut{Output}{Output}
\Input{\newline - Graph $G_t =(V, E_t)$
       \newline - Prior vectors $\phi_{(t-1)}$
       \newline - Probabilities $\alpha_\mathbf{v}$ and $\beta_{{v}, {u}}$ 
       }
\Output{Retrofitted vectors $\phi$}
       
$\phi \leftarrow \phi_{(t-1)}$  \tcp{initialization}
\Repeat {convergence}{
\For {all ${v} \in V_t$} { 
	$\phi_{t}({v}) \leftarrow \frac{\alpha_{v} \phi_{(t-1)} ({v}) +  \sum_{{u}} \beta_{{v}, {u}}  \phi_{t} ({u})}{\alpha_{v} + \sum_{{u}} \beta_{{v}, 
{u}}}$
}
} 
\caption{Jacobi method for retrofitting.}
\label{alg:ret}
\end{algorithm}

In our experiments, we set $\beta_{{v}, {u}} = \frac{1}{degree({v})}$, and use the same $\alpha$, which we tune using a held-out validation set, for all nodes $v \in V$. In other words, we vary weights for \emph{temporal smoothness} while fixing the weights for \emph{network proximity}. It is clear from Eq. \ref{alg:ret} that $\phi_t(v)$ is a convex combination
of $v$'s embedding at $t-1$ and the centroid of $u$'s neighbors' embeddings at $t$.
Algorithm \ref{alg:ret} formally describes the training procedure of our retrofitted model. We experiment with several static embedding models to learn the embeddings of the first snapshot, $\phi_1$ (see Section \ref{sec:models_compared} for details).  

Except for the first time-stamp, the retrofitted model does not ``learn'' from data about how to transform the embeddings, rather it presumes smoothing criteria. This presumption is effective when the smoothing criteria are met, but may be ineffective otherwise. Therefore, retrofitting is not a generic solution. In addition, retrofitting is limited because it requires a network snapshot $G_t$ to perform inference at time $t$.
To address these limitations, we propose linear transformation models.   





\subsection{Linear Transformation models} \label{sec:lp}

In a dynamic network, we can expect that the network evolves by following a domain dependent pattern. So, the vertex representation vectors of two different time-stamps should have a similar transformation. In our transformation based models, we exploit this similarity by learning a linear mapping from a source ($\phi_{t-1}$) to a target ($\phi_{t}$) embedding space.

Our goal is to learn a transformation matrix, $W \in \real^{d \times d}$, that can transform $\phi_{t-1}$ to $\phi_{t}$, given the network snapshots, $\Gs = (G_1, G_2,$ $\ldots, G_{T})$, and their corresponding statically-learned network embedding matrices, $\Phi = (\phi_1, \phi_2, \ldots, \phi_{T})$. We explore two types of models: homogeneous and heterogeneous. In the homogeneous model, we assume that the transformation matrix is the same for the time-stamps, $1$ to $T$, i.e., $W$ is shared across snapshots. On the other hand, for the heterogeneous
model, we assume a different $W$ for each pair of time-stamps, resulting in $T-1$ different transformation matrices. 
We form the final transformation matrix, $W$, by combining the $T - 1$  matrices.

\noindent\textbf{Homogeneous Transformation Model}\newline
Our homogeneous transformation model is shown in Figure \ref{fig:models}(b). First, we construct both a source matrix $X$ by vertically stacking the embedding matrices $\phi_1$, $\phi_2$, \ldots, $\phi_{T-1}$, and a target matrix $Z$ by vertically stacking the matrices $\phi_2$, $\phi_3$, \ldots, $\phi_{T}$ (as shown in Eq. \ref{eq:linproj}), given the sequence of (static) embedding matrices $\Phi = (\phi_1, \phi_2, \ldots, \phi_{T})$.
Corresponding rows $X_u$ and $Z_u$ represent the embedding vectors for node $u$ at network snapshots $G_t$ and $G_{t+1}$, respectively, for $t = 1, 2, \ldots, T-1 $. To learn the matrix $W$, we minimize the objective function: 

\begin{equation}
J(W) = ||WX - Z ||^{2}, \hspace{0.1in} \text{where} \hspace{0.1in} X = \begin{bmatrix}
    \phi_{1} \\
    \phi_{2} \\
    \vdots \\
    \phi_{T-1}  
\end{bmatrix} ;
Z = \begin{bmatrix}
    \phi_{2} \\
    \phi_{3} \\
    \vdots \\
    \phi_{T}
\end{bmatrix}.
\label{eq:linproj}
\end{equation}

We solve Eq. \ref{eq:linproj} with gradient descent. One can use stochastic gradient descent with minibatch to scale to large $X$ and $Z$ matrices. 
 





\noindent\textbf{Heterogeneous Transformation Model} \newline
In our heterogeneous model, we minimize an  objective function similar to Eq.~\ref{eq:linproj}, but learn a different projection matrix for each pair of network snapshots.  Given $T$ different embedding matrices $\Phi = (\phi_1, \phi_2, \ldots, \phi_{T})$, we learn $T-1$ different transformation matrices by minimizing the objective function:
\begin{equation}
J(W_{t}) = || W_t \phi_t - \phi_{t+1} ||^{2}, \hspace{0.1in} for~~ t = 1, 2, \ldots, (T-1).
\label{eq:linproj3}
\end{equation}

Then, we obtain the final transformation matrix $W$ by combining the projection matrices from times $t = 1, 2, \ldots, (T-1)$. Figure \ref{fig:models}(c) depicts this process for three snapshots.
To smooth the projection matrices from Eq.~\ref{eq:linproj3}, we experiment with different smoothing combinations:
\begin{enumerate}[(a)]
\item \textit{Uniform smoothing:} We weight all projection matrices equally, and linearly combine them: 
\begin{align}
(avg) \quad W = \frac{1}{T-1} \sum_{t=1}^{T-1} W_t \label{eq:avg}.
\end{align}

\item \textit{Linear smoothing:} We increment the weights of the projection matrices linearly with time: 
\begin{equation}
(linear) \quad W = \sum_{t=1}^{T-1} \frac{t}{T-1} W_t. \label{eq:linear_smoothing}
\end{equation}

\item \textit{Exponential smoothing:} We  increase weights exponentially, using an exponential operator (exp) and a weighted-collapsed tensor (wct):
\begin{align}
(exp) \quad W &= \sum_{t=1}^{T-1} \exp^{\frac{t}{T-1}} W_t \label{eq:exp_smoothing}\\
(wct) \quad W &= \sum_{t=1}^{T-1} (1 - \theta)^{T-1-t} W_t. \label{eq:weighted_collasped_tensor}
\end{align}
\end{enumerate}

\section{Experimental Settings}
\label{sec:eval_tasks}
To evaluate the performance of our dynamic network embedding models visually, we show a
network visualization video over time using the Enron email dataset (see section~\ref{sec:vis}). For quantitative evaluation,
we solve the \textbf{temporal link prediction} task using the vertex embedding vectors from our models. Temporal link prediction is an extension of the well-known missing link prediction problem in a static network. It is defined as follows. Given a sequence of $T$ snapshots of an evolving network, $\Gs = (G_1, G_2, \ldots, G_{T})$, predict the links in $G_{T+1}$; in other words, construct a function $f(u, v)$ that predicts whether an edge $e(u, v)$ exists between any two nodes $u, v \in V_{T+1}$. 
In our retrofitted model, we use Hadamard product (element-wise multiplication) of node embeddings of time $T$ (i.e., $\phi_T(u)$ and $\phi_T(v)$) as the input feature representation of the node-pair $\{u, v\}$. For the transformation models, 
we use the hadamard product of node embeddings of time $T+1$ i.e., $\phi_{T+1}(u)$ and $\phi_{T+1}(v)$ learned using $W\cdot\phi_T$ as the input feature representation. Then, we use a logistic regression model (a Scikit-learn implementation with default parameter settings) to obtain $f$.

For each dataset described in Section \ref{sec:dataset}, we randomly select 50\% of the total positive edges from $G_T$ as training and 30\% as test, and we leave the remaining as validation. 
We randomly select equal amounts of negative edges for the train, test and validation sets for a balanced 
classification.
We tune the hyper-parameters on the validation set, and evaluate our models on the test sets. 
We repeat this process 10 times to get 10 different negative edge sets; and we report our average performance over these sets. 

\subsection{Datasets} \label{sec:dataset}
We perform temporal link prediction on nine datasets of three classes of networks: four academic collaboration networks, three
messaging networks, and two social networks. The datasets vary in size and density. In Table \ref{tab:dataset}, we report the number of nodes, the number of
distinct edges (across all time stamps), and the number of interactions (counting plurality of an edge across different
timestamps) for each dataset. We have published all the processed datasets along with our code release.

\begin{table}
\centering
\ra{0.9}
\resizebox{0.95\linewidth}{!}{%
\setlength\tabcolsep{8pt}
\begin{tabular}{lllll}
\toprule
{Data}& {\#Nodes} & {\#Edges} & Interactions & {\#Snapshot} \\
\midrule
\sc{DBLP2} & 315 & 943 & 2552 & 10 \\
\sc{DBLP3} & 653 & 3379 & 9080 &  10 \\
\sc{NIPS}  & 2865 & 4733 & 5461 &  17 \\
\sc{HepPH} & 28093 & 3148447 & 3718015 & 9 \\\hline 
\sc{CollegeMsg} & 1899 &13838 &18127  & 10 \\
\sc{SMS-A} & 44430&52222 &144164& 30 \\
\sc{Email-EU} & 986&16064&81147& 30\\\hline 
\sc{Facebook}  &63731 & 817035&817035& 5 \\
\sc{Facebook2} &663 &5271&11697& 9 \\
\bottomrule
\end{tabular}
}
\vspace{0.1in}
\caption{Temporal link prediction datasets. Nodes and Edges denote the 
distinct number of vertices and edges over all the time-stamps. We also
report the number of distinct interactions after removing self-edges. 
Number of snapshots denote 
the total number of time spans of the data.}
\label{tab:dataset}
\vspace{-0.2in}
\end{table}

\noindent \textbf{(Collaboration Networks) \sc{DBLP2, DBLP3, NIPS, HepPH:}} 
Both DBLP2 and DBLP3 datasets (obtained from \url{arnetminer.org}) have $10$ time stamps with the paper citation information of about $49,455$ author-pairs and around $1.4$ million papers. 
To create the co-authorship network, we add edges between people who co-authored a paper.
We consider publications between $2000$-$2009$, each year as a time stamp. 
Since DBLP2 and DBLP3 datasets are very sparse, we preprocess the data to retain only the active authors, whose last published papers are on or after the year $2010$. For DBLP2,
we retain authors who participated in at least two publications in seven or more time 
stamps. For DBLP3, we retain the authors with at least four publications in seven or more time-stamps.
The NIPS \cite{globerson2007euclidean} dataset consists of collaboration information among 2865 NIPS authors. The dataset contains 17 snapshots for volumes 1-17. HepPH is the collaboration graph of authors of scientific papers from arXiv's High Energy Physics - Phenomenology. An edge between two authors represents a coauthored publication, and its time-stamp denotes the publication date. We divide HepPH dataset into nine snapshots.

\noindent \textbf{(Messaging Networks) \sc{CollegeMsg, Email-EU, SMS-A \cite{paranjape2017motifs}:}} The CollegeMsg dataset is comprised of messages from an online social network at the University of California, Irvine. An edge represents a private message between users.
Email-EU dataset is a collection of emails between members of a European research institution, such that an edge represents an email.
In the SMS-A dataset, an edge is an SMS text between persons.
We divide CollegeMsg into 10 shapshots, and the other two datasets into 30.

\noindent \textbf{(Social Networks) \sc{Facebook, Facebook2:}}
In the {\sc Facebook} dataset, a node represents a user in 
the Facebook friendship network, and an edge represents a friendship relation between two users. Time-stamps denote the time the friendship was established. We divide this dataset into 5 snapshots. 
{\sc Facebook2} is a network of Facebook wall posts \cite{viswanath2009evolution}. Each node is a Facebook user account, and each edge represents a user's post on another user's wall. {\sc Facebook2} has $9$ time-stamps because we preprocess this dataset in the same way as Xu
\cite{KevinS:2015}, where each time-stamp represents 90 days of wall posts.

\subsection{Evaluation Metrics}

To evaluate link prediction performance, we use three metrics:  area under the Receiver Operating Characteristic (ROC) curve (AUC), area under the Precision-Recall curve (AUPRC) \cite{Davis:2006}, and Normalized Discounted Cumulative Gain (NDCG), an information retrieval metric. AUC is equal to the probability that a classifier will, for the link prediction task, rank a randomly chosen positive instance (a node-pair which has an edge at time $T$) higher than a randomly chosen negative instance (a node-pairs with no edge at time $T$). AUC values range from 0.0 to 1.0.
The second metric AUPRC considers 
the ranked sequence of node pairs based on their likelihood to form an edge at time $T$. 
We create a precision-recall curve by computing precision and recall at every position in the ranked sequence of node pairs. AUPRC is the average value of precision over the interval of lowest recall (0.0) to highest recall (1.0)
AUPRC values range from 0.0 to 1.0. 
NDCG measures the performance of a link prediction system based on the graded relevance of recommended links. $NDCG_P$ varies from 0.0 to 1.0, where 1.0 represents the ideal ranking of edges. 
$P$, a number chosen by the user, is the number of links ranked by the method.
We choose $P=50$ in all our experiments. 

\subsection{Competing Methods}

Our objective is to compare the relative quality of latent vertex embeddings in a dynamic network. So, we only compare our proposed methods with existing vertex embedding models which can provide explicit latent vectors for all the vertices at every time-stamp of a dynamic network. The majority of dynamic link prediction methods~\cite{Gunes2015,Rahman.Hasan:16,Li:2014,Dunlavy:2011,KevinS:2015} do not satisfy this requirement, except Zhu et al.
~\cite{Zhu.Guo:16}. They propose a method for performing dynamic link prediction called BCGD (Block Co-ordinate Gradient Descent) which performs non-negative matrix factorization with temporal smoothness. From the factorization of the adjacency matrix at each time stamp, we can obtain latent vectors of the vertices. BCGD  minimizes the objective function:
\begin{align}
J_{BCGD} & =  \underbrace{\sum_{t=1}^{T-1} || G_t - \phi_t(u)\phi_t(v) ||^{2}}_\text{Network Proximity} \nonumber \\ 
	     & +   \underbrace{\lambda \sum_{t=1}^{T-1} \sum_{u} 1 - \phi_t(u) {\phi_{t-1}(u)}^T}_\text{Temporal Smoothing} s.t. \hspace{0.05in} \phi_t \geq 0 \label{eq:bcgd}. 
\end{align}
The temporal smoothing part of Equation \ref{eq:bcgd} penalizes sharp change of latent position of a node $u$, whereas, the first part captures the latent proximity. 
We use the author-provided implementation of the incremental-BCGD algorithm and tune its
parameters identically along with our proposed methods. 

\subsection{Different Configurations of the Proposed Models} \label{sec:models_compared}
We vary our three proposed models: Homogeneous (homogeneous transformation),
Heterogeneous (heterogeneous transformation), and Retrofitted models, by choosing seven different base representation-learning methods: three random-walk based (Deepwalk, LINE, Node2Vec), two matrix factorization based (PCA, tsvd), and two manifold based (LLE, ISOMAP).
For the transformation models we utilize the base method to learn the vertex representation vectors on the snapshots 1 to (T-1). Our retrofitted models use
the base representation methods to learn the vertex representation vectors on the first snapshot of the graph. 
To vary our heterogeneous transformation models further, we experiment with different smoothing functions: Uniform,  Linear, and Exponential (exp and wct). For fair comparison, we set the representation dimensions equal to 64 for all models. However, we have reported results over other representation dimensions (see Section \ref{sec:lat-dim}).

DeepWalk, LINE, and Node2Vec methods are trained with stochastic gradient descent. We used negative sampling, with 5 noise samples,
to significantly decrease training time. We also used subsampling of frequent words. We tune DeepWalk's window size parameter from the set \{8, 12, 15\}. We tune LINE's iteration parameter from the set \{5, 10, 20\} (millions of iterations). We tune Node2Vec's $p$ and $q$ parameters, which control the amount of exploration vs exploitation in the random walk, from the set \{0.1, 0.3, 0.5\}. For PCA, tsvd, LLE, and IsoMap, we use the implementation provided by scikit-learn with the default settings.

In our Retrofitting models, we iterate 20 times. We tune $\alpha_v$, which controls the weight of the prior vector, from the set of values: \{0.1, 1, 10\}.  For our Homogeneous and Heterogeneous models, we use Batch Gradient Descent with 10,000 iterations. We also perform gradient clipping, which clips values of multiple tensors by the ratio of the sum of their norms, with a clipping ratio of 5.0. 


\begin{table*} [!ht]
\begin{subtable}{1.0\linewidth}
\centering
\resizebox{0.95\linewidth}{!}{%
\begin{tabular}{llll|llll}
\toprule
	\multicolumn{4}{c}{DBLP2} & \multicolumn{4}{c}{DBLP3} \\ 
{Method}& {AUC $\pm$ sd} & {AUPRC $\pm$ sd} & $NDCG_P$ $\pm$ sd  & Method & {AUC $\pm$ sd} & {AUPRC $\pm$ sd} & $NDCG_P$ $\pm$ sd \\\hline 
BCGD & 0.7932 $\pm$ 0.03 & 0.8023 $\pm$ 0.03 & 0.8686 $\pm$ 0.02 & BCGD & 0.8564 $\pm$ 0.01 & 0.8599 $\pm$ 0.02 & 0.9524 $\pm$ 0.02 \\
RET (tsvd) & \textbf{0.8360 $\pm$ 0.02} & \textbf{0.8571 $\pm$ 0.01} & \textbf{0.9317 $\pm$ 0.02} & RET (LLE) & \textbf{0.8898 $\pm$ 0.01} & \textbf{0.8926 $\pm$ 0.01} & \textbf{0.9776 $\pm$ 0.01} \\ 
HomoLT (DeepWalk) & 0.7422 $\pm$ 0.03 & 0.7691 $\pm$ 0.04 & 0.8496 $\pm$ 0.05 & HomoLT (LINE) & 0.7557 $\pm$ 0.02 & 0.7818 $\pm$ 0.02 & 0.9671 $\pm$ 0.01 \\
HeterLT (DeepWalk, avg) & 0.7413 $\pm$ 0.02 & 0.7798 $\pm$ 0.02 & 0.8662 $\pm$ 0.02 & HeterLT (LINE, linear) & 0.8301 $\pm$ 0.01 & 0.8680 $\pm$ 0.007 & 0.9934 $\pm$ 0.004 \\
\bottomrule
\end{tabular}
}
\caption{DBLP2 and DBLP3}
\label{tab:result-citation1}
\end{subtable}

\begin{subtable}{1.0\linewidth}
\centering
\resizebox{0.95\linewidth}{!}{%
\begin{tabular}{llll |llll}
\toprule
\multicolumn{4}{c}{HepPH} &\multicolumn{4}{c}{NIPS} \\
{Method}& {AUC $\pm$ sd} & {AUPRC $\pm$ sd} & $NDCG_P$ $\pm$ sd & {Method}& {AUC $\pm$ sd} & {AUPRC $\pm$ sd} & $NDCG_P$ $\pm$ sd \\\hline 
BCGD                & 0.5732 $\pm$ 0.003 & 0.6170 $\pm$ 0.003 & 0.5542 $\pm$ 0.02 & BCGD	& 0.5157 $\pm$ 0.002	& 0.5457 $\pm$ 0.02 &	0.6118 $\pm$ 0.02 \\
RET (LINE)     & 0.5677 $\pm$ 0.004 & 0.5782 $\pm$ 0.003  & 0.8659 $\pm$ 0.04 & RET (LLE) &	0.5427 $\pm$ 0.04	& 0.5542	$\pm$ 0.02 &  0.5988 $\pm$ 0.08 \\
HomoLT (DeepWalk) & 0.5889 $\pm$ 0.003 &  0.6269 $\pm$ 0.005 & 0.8376 $\pm$ 0.04 & HomoLT (Node2Vec) & \textbf{0.5633 $\pm$ 0.01} & \textbf{0.6123 $\pm$ 0.01} & 0.7947 $\pm$ 0.02 \\
HeterLT (DeepWalk, avg)& \textbf{0.6058 $\pm$ 0.02} & \textbf{0.6346 $\pm$ 0.02} &\textbf{0.9300$\pm$ 0.03} & HeterLT (Node2Vec, wct)  & 0.5581 $\pm$ 0.01 & 0.6211 $\pm$ 0.01 & \textbf{0.8072 $\pm$ 0.03} \\
\bottomrule
\end{tabular}
}
\caption{HepPH and NIPS}
\label{tab:tab2}
\end{subtable}

\begin{subtable}{1.0\linewidth}
\centering
\resizebox{0.95\linewidth}{!}{%
\setlength\tabcolsep{8pt}
\begin{tabular}{llll|llll}
\toprule
\multicolumn{4}{c}{Email-EU} &\multicolumn{4}{c}{CollegeMsg} \\
{Method}& {AUC $\pm$ sd} & {AUPRC $\pm$ sd} & $NDCG_P$ $\pm$ sd & {Method}& {AUC $\pm$ sd} & {AUPRC $\pm$ sd} & $NDCG_P$ $\pm$ sd \\\hline 
BCGD                 & 0.6215 $\pm$ 0.01 & 0.5946 $\pm$ 0.02 & 0.6154 $\pm$ 0.13 & BCGD                & 0.6663 $\pm$ 0.01 &  0.6691 $\pm$ 0.02 &  0.7266 $\pm$ 0.06 \\
RET (LINE)  & 0.9049 $\pm$ 0.005 &  0.9009 $\pm$ 0.009 & 0.9725 $\pm$ 0.02 & RET (PCA) & 0.6291 $\pm$ 0.01 & 0.6435 $\pm$ 0.02 & 0.8381 $\pm$ 0.06 \\ 
HomoLT (LINE)  & 0.8789 $\pm$ 0.009  & 0.8694 $\pm$ 0.01 & 0.9705 $\pm$ 0.01 & HomoLT (LINE) &  {0.7460 $\pm$ 0.01} & {0.7788  $\pm$ 0.01}  &  {0.9571 $\pm$ 0.01} \\
HeterLT(LINE, wct)  & \textbf{0.9211 $\pm$ 0.008} & \textbf{0.9283 $\pm$ 0.006} & \textbf{0.9923 $\pm$ 0.008} & HeterLT (LINE, wct)    &  \textbf{0.7517 $\pm$ 0.01} &  \textbf{0.7913 $\pm$ 0.008}  & \textbf{0.9685 $\pm$ 0.01} \\
\bottomrule
\end{tabular}
}
\caption{Email-EU and CollegeMsg}
\label{tab:tab3}
\end{subtable}

\begin{subtable} {1.0\linewidth}
\centering
\resizebox{0.95\linewidth}{!}{%
\begin{tabular}{llll|llll}
\toprule
\multicolumn{4}{c}{SMS-A} &\multicolumn{4}{c}{SMS-A} \\
{Method}& {AUC $\pm$ sd} & {AUPRC $\pm$ sd} & $NDCG_P$ $\pm$ sd  & {Method}& {AUC $\pm$ sd} & {AUPRC $\pm$ sd} & $NDCG_P$ $\pm$ sd \\\hline 
BCGD                & 0.7350 $\pm$ 0.003 &  0.7770 $\pm$ 0.003 &  0.9312 $\pm$ 0.01  & HomoLT (DeepWalk) & 0.6306 $\pm$ 0.005 & 0.6861 $\pm$ 0.006 & 0.9850 $\pm$ 0.01 \\
RET (Node2Vec)         & \textbf{0.7737 $\pm$ 0.007} &  \textbf{0.8089 $\pm$ 0.006}  &  \textbf{1.00 $\pm$ 0.00} & HeterLT (DeepWalk, avg)  & 0.6413 $\pm$ 0.03 & 0.6969 $\pm$ 0.02 & 0.99 $\pm$ 0.03 \\
\bottomrule
\end{tabular}
}
\caption{SMS-A}
\label{tab:tab4}
\end{subtable}

\begin{subtable}{1.0\linewidth}
\centering
\resizebox{0.95\linewidth}{!}{%
\setlength\tabcolsep{8pt}
\begin{tabular}{llll| llll}
\toprule
\multicolumn{4}{c}{Facebook} &\multicolumn{4}{c}{Facebook2} \\
{Method}& {AUC $\pm$ sd} & {AUPRC $\pm$ sd} & $NDCG_P$ $\pm$ sd & {Method}& {AUC $\pm$ sd} & {AUPRC $\pm$ sd} & $NDCG_P$ $\pm$ sd \\\hline 
BCGD & 0.6431 $\pm$ 0.002 & 0.6576 $\pm$ 0.003 & 0.3694 $\pm$ 0.02  & BCGD & 0.7537 $\pm$ 0.02 & 0.7190 $\pm$ 0.02  & 0.7957 $\pm$ 0.05 \\
RET (Node2Vec) & \textbf{0.859 $\pm$ 0.004} & \textbf{0.859 $\pm$ 0.005} & 0.9817 $\pm$ 0.006 & RET(PCA) & \textbf{0.8202 $\pm$ 0.01} & \textbf{0.8144 $\pm$ 0.01} & \textbf{0.9519 $\pm$ 0.02} \\
HomoLT (DeepWalk) & 0.6061 $\pm$ 0.004 & 0.6141 $\pm$ 0.004 & 0.7113 $\pm$ 0.02 & HomoLT (DeepWalk) & 0.7252 $\pm$ 0.02 &0.7144 $\pm$ 0.02  & 0.8422 $\pm$ 0.03\\
HeterLT (LINE, avg)  & 0.6258 $\pm$ 0.02 & 0.6983 $\pm$ 0.02 & \textbf{0.9823 $\pm$ 0.03} & HeterLT (Node2Vec, linear) & 0.7792 $\pm$ 0.01 & 0.7788 $\pm$ 0.01 & 0.9200 $\pm$ 0.01\\
\bottomrule
\end{tabular}
}
\caption{Facebook and Facebook2}
\label{tab:tab5}
\end{subtable}
\caption{Performance of seven of our Homogeneous models, 28 of our Heterogeneous models, and 7 of our retrofitted models, on nine datasets. The highlighted results are statistically significant over the baseline with p<0.001. For fair comparison, we set the latent dimension size to 64. }
\label{tab:all}
\vspace{-0.3in}
\end{table*}

\section{Results and Discussion}
\label{sec:tempn2v:expresult}

In Table~\ref{tab:all}, we show the comparison between the competing method (BCGD) and
our proposed models: retrofitting (RET), homogeneous transformation (HomoLT),
and heterogeneous transformation (HeterLT). 
Our models' performances vary for each base embedding method they implement. Therefore, we only report the results from the best-performing base embedding model (name in parentheses).
In cases where the best-performing base models differ for different metrics, we present the results of the base model that is best overall. For all datasets
and all metrics the best performing model's performance is shown in boldface font.

Table~\ref{tab:all}. shows that one of our three proposed methods outperforms BCGD in all three metrics over all datasets.
This suggests that the latent embedding vectors from our proposed models
are better for link prediction than BCGD's embedding vectors.
We think BCGD may under-perform because it can only exploit edge-based proximity when learning latent embedding vectors by factoring the adjacency matrix. Whereas, our models capture more complex network proximity by implementing, as base embedding models, state-of-the-art static node embedding methods.
In addition, our local and global temporal smoothness methods provide a better dynamic network model than BCGD. Below, We present the results in detail by grouping them over the three different kinds of networks.

\begin{figure*}[!ht]\centering
   \centering
    \begin{subfigure}{0.33\linewidth}
        \centering
        \includegraphics[scale=0.42]{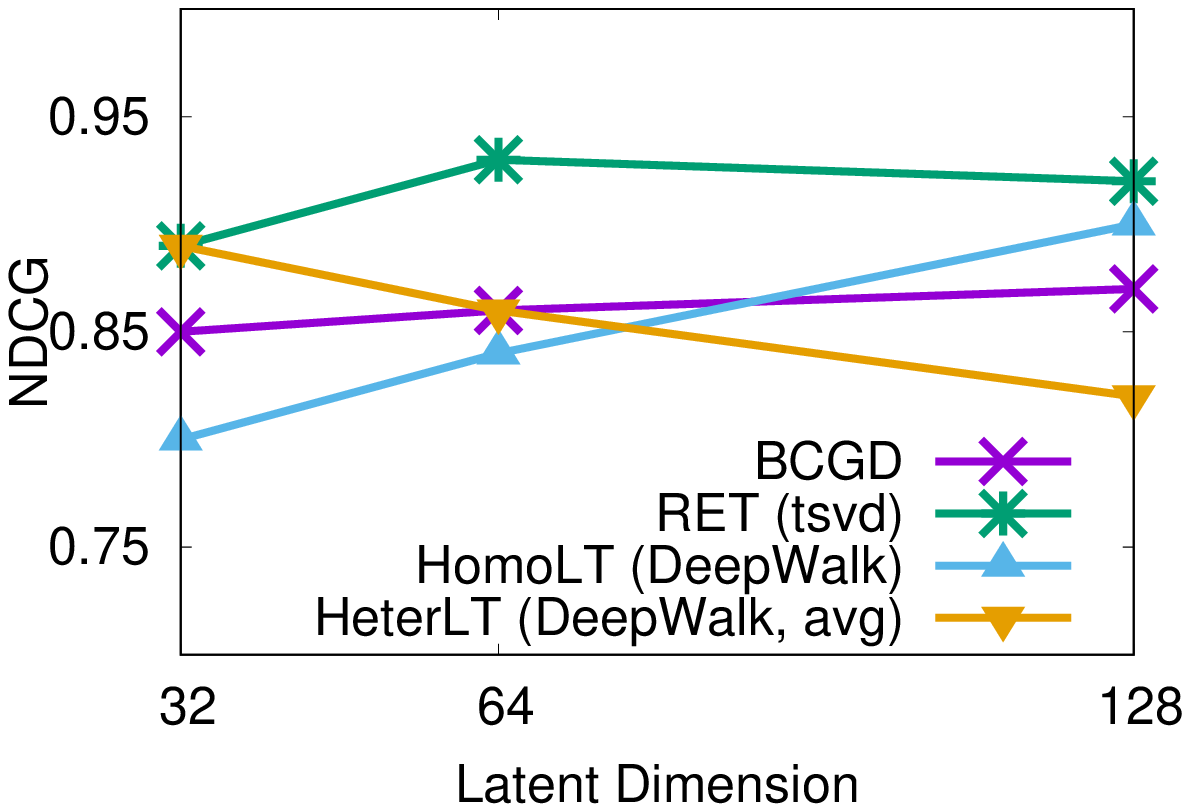} 
        \caption{DBLP2} \label{fig:a}
    \end{subfigure}
    \begin{subfigure}{0.33\linewidth}
        \centering
        \includegraphics[scale=0.42]{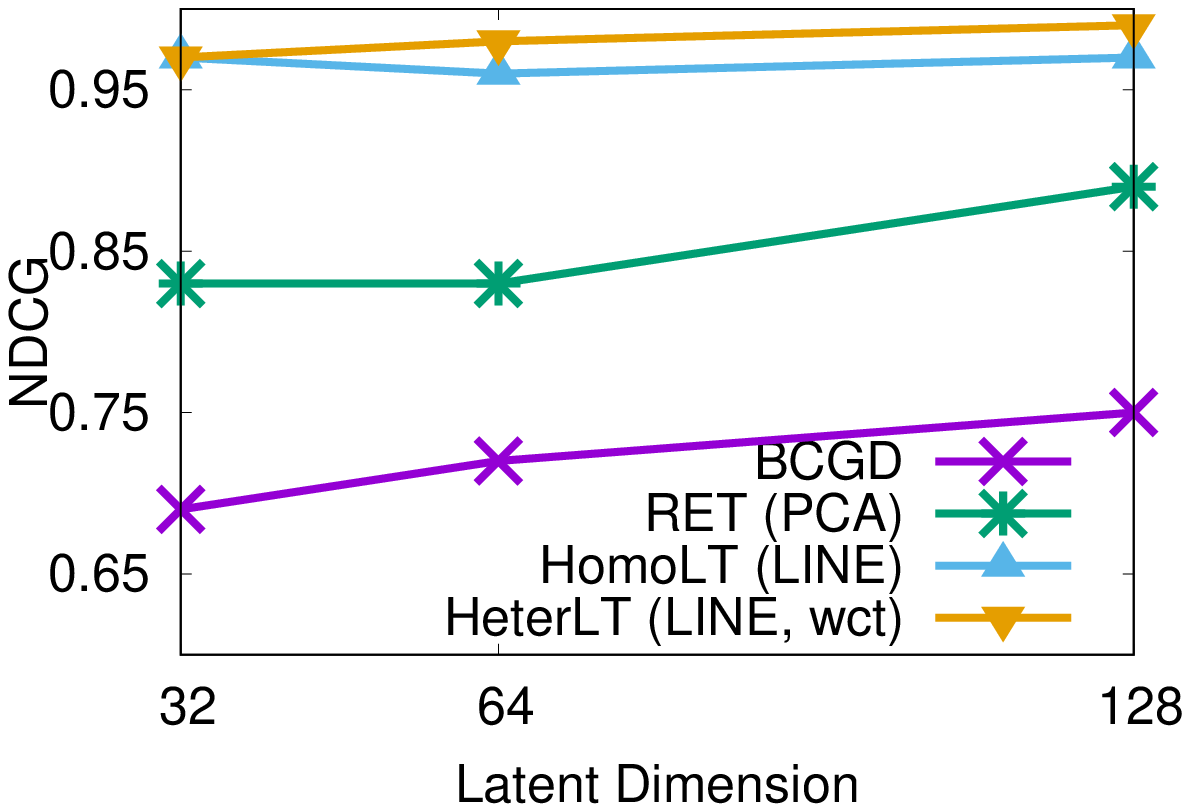} 
        \caption{CollegeMsg} \label{fig:b}
    \end{subfigure}
   \begin{subfigure}{0.33\linewidth}
       \centering
       \includegraphics[scale=0.42]{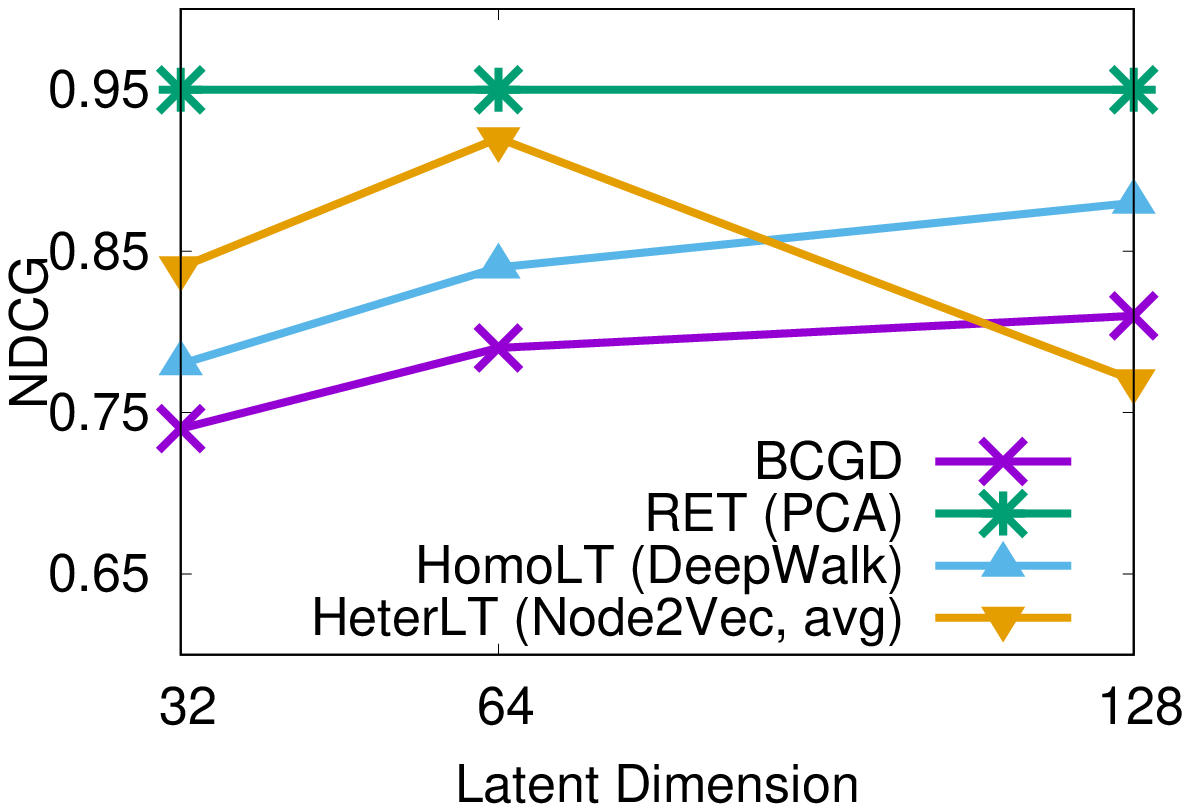} 
       \caption{Facebook2} \label{fig:c}
   \end{subfigure}
\vspace{-0.2in} 
\caption{Effect on Latent Dimension. We evaluate the performance of our best models along with the baseline model to study the effect of different latent dimensions (in our case, 32, 64, and 128).}
\label{fig:latentDimension}
\end{figure*}

\subsection{Link Prediction in Citation Network}
Results of three collaboration networks are given in Table~\ref{tab:result-citation1} and ~\ref{tab:tab2}.
Among the citation networks, DBLP2 and DBLP3 results are similar (see Table ~\ref{tab:result-citation1}); for both the datasets, the
retrofitted model (RET) performs the best in all three metrics. For DBLP2, RET improves
the AUC, PRAUC, and NDCG values of the competing BCGD method by 0.04, 0.06, and 0.06 units, which translates to 
5\%, 7\%, and 7\% improvement, respectively. For DBLP3, the improvement of RET on these three
metrics are 0.04, 0.03, and 0.02. Interestingly, the base learning models of the best performing
RET differs over these datasets, for DBLP2, it is tsvd and for DBLP3, it is LLE. The performance
of HomoLT and HeterLT in comparison to BCGD are mixed over different metrics. For instance, 
they are better in NDCG metric, but marginally worse in the AUC, and PRAUC metrics. 

For the NIPS dataset (see Table \ref{tab:tab2}), the transformation models (HomoLT and HeterLT) are the
best performing models. Homogeneous and Heterogeneous models with Node2Vec as the base embedding model
have the best performance; HomoLT is the winner in AUC and PRAUC, and HeterLT is the winner in NDCG.
In fact, the NDCG value of HeterLT is .8072, which is better than the same for BCGD by 0.19 units, more than
30\% improvement! For AUC metric, the improvement is around 0.04 unit, and for PRAUC metric the improvement
is around 0.06 unit for both the homogeneous and heterogeneous models. 
In NIPS dataset, RET model has a mixed performance compared to BCGD, the former wins in AUC and PRAUC, but loses in NDCG, both marginally.
An explanation of sub-optimal performance by the retrofitted model in this dataset may be because of its large number of time snapshots (17); because of this, the vectors of the last time snapshot, which are obtained by 16 iterations of retrofitting of the base embedding vectors of first snapshot, 
may have wandered away from their optimal position. Another explanation is that in this dataset the number of unique edges (4733) is quite close to the number
of total interaction (5461); i.e., edges are not repeated so each new snapshot is very different than the previous snapshots and retrofitting may not
the ideal approach for capturing the temporal smoothness of this dataset. HepPH dataset also has the same behavior as NIPS (results are available in the same table). In this dataset the best
performing model is HeterLT with Deepwalk as the base embedding. In fact, for
the HepPH dataset, HeterLT has more than 30\% improvement over the BCGD model in NDCG metric, and
around 5\% improvement in two other metrics. For this dataset also, RET has mixed performance with respect to BCGD. The suboptimal performance of retrofitting
models may be due to the very small ratio of the number of distinct edges and the total number of interactions.

\subsection{Link Prediction in Messaging Network}
The results of the messaging datasets are shown in Table~\ref{tab:tab3} and Table~\ref{tab:tab4}.

For Email-EU, all of our proposed methods perform better than the BCGD model by a substantial margin; HeterLT (LINE) performs the best in all three metrics combined. For example, HeterLT (LINE) model improves the NDCG value of BCGD by 0.38 units, from 0.6154 to 0.9959! Similar large improvements can also be seen in the other two metrics (please see Table~\ref{tab:tab3} (Left) for the detailed results).
For this dataset, retrofitting models also perform substantially better than BCGD. A possible explanation is the high ratio of
distinct edge and the number of interactions, i.e., the earlier edges are repeated in later iterations, so the retrofitting based temporal smoothness of the node vectors are sufficient for capturing the network dynamics
in this dataset. CollegeMsg dataset also has similar behavior with HeterLT (LINE) as the winner among all, again with substantial performance gain (around 20\% to 30\% improvement of performance value across all three
metrics). For this dataset, retrofitting results are poor which could be due to small ratio of distinct edge vs interaction count, and large number of temporal snapshots.
For SMS-A dataset (results is shown on both side of Table~\ref{tab:tab4}), retrofitted
method with Node2Vec as the initial representation generator performs the best. The model achieves 0.04 unit improvement in AUC, 0.03 unit in AUPRC and 0.07 unit in NDCG over BCGD. For this dataset, the ratio of distinct
edge vs interaction count is higher than the CollegeMsg dataset, which could be a reason for the RET model to perform better.


\subsection{Link Prediction in Social Network}
Results on social networks are shown in Table~\ref{tab:tab5}.
For Facebook and Facebook2, the retrofitted
method with Node2Vec and PCA performs better than BCGD. RET (Node2Vec) performs the best in Facebook dataset. The model gains 0.21 unit improvement over BCGD in AUC metric, 0.20 unit in AUPRC, and around 0.60 unit in NDCG metric. For Facebook2 dataset, RET (PCA) performs the best. RET (PCA) improves 0.07 unit over BCGD in AUC metric, 0.10 unit in AUPRC metric, and 0.16 unit in the NDCG metric. Heterogeneous method with Node2Vec as the representation generator along with linear or exponential smoothing operator also performs better than BCGD achieving around 0.06 unit improvement in AUPRC, and 0.13 unit improvement in NDCG metric. The Facebook datasets 
have small number of timestamps, which is a likely reason for RET model to perform better than the transformation based models on these datasets.

\begin{figure*}[!h]
\begin{subfigure}{0.3\linewidth}
\includegraphics[scale=0.27]{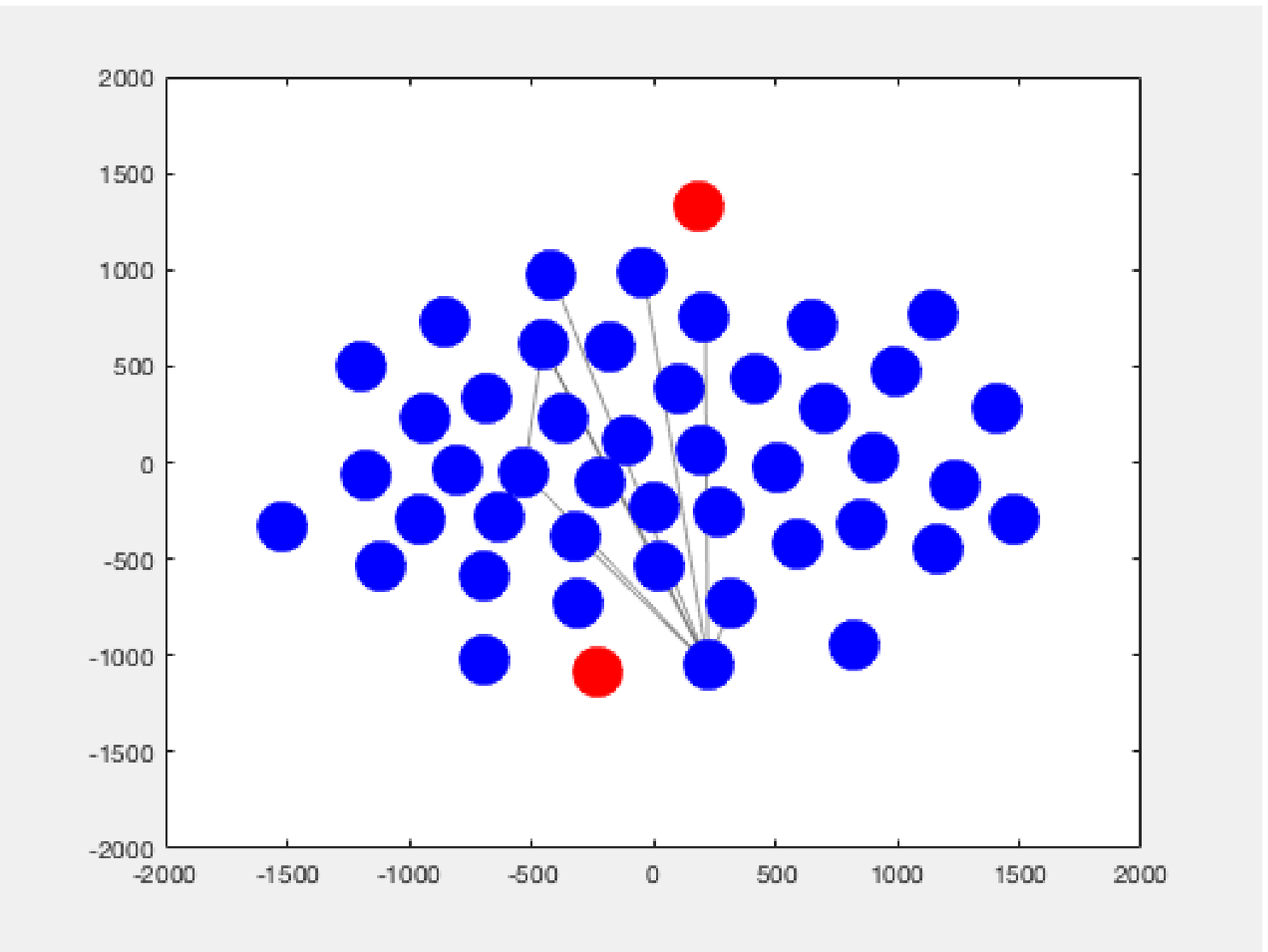}
\caption{Snapshot 1}
\end{subfigure}
\hspace{0.1in}
\begin{subfigure}{0.3\linewidth}
\includegraphics[scale=0.27]{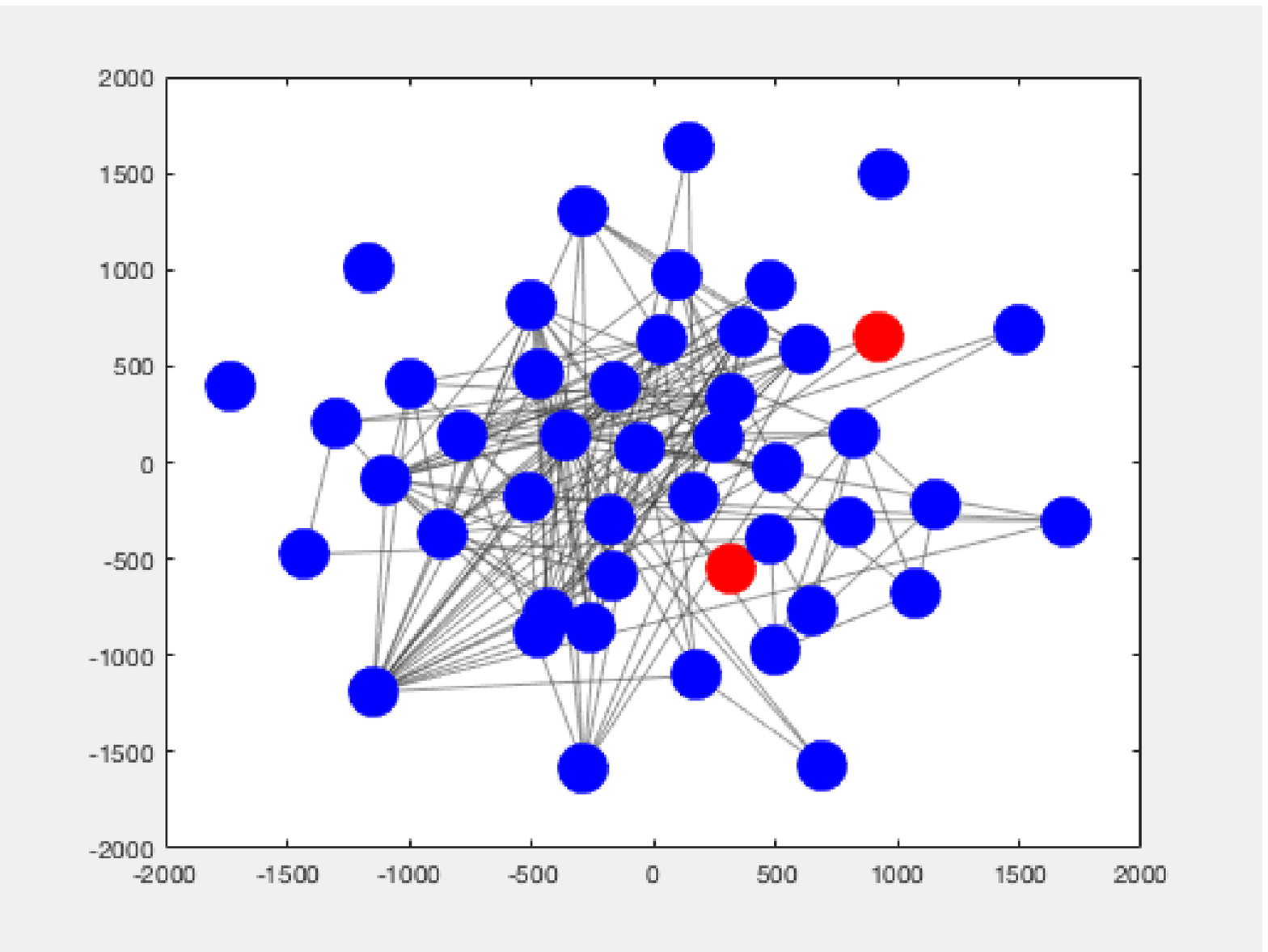}
\caption{Snapshot 11}
\end{subfigure}
\hspace{0.1in}
\begin{subfigure}{0.3\linewidth}
\includegraphics[scale=0.27]{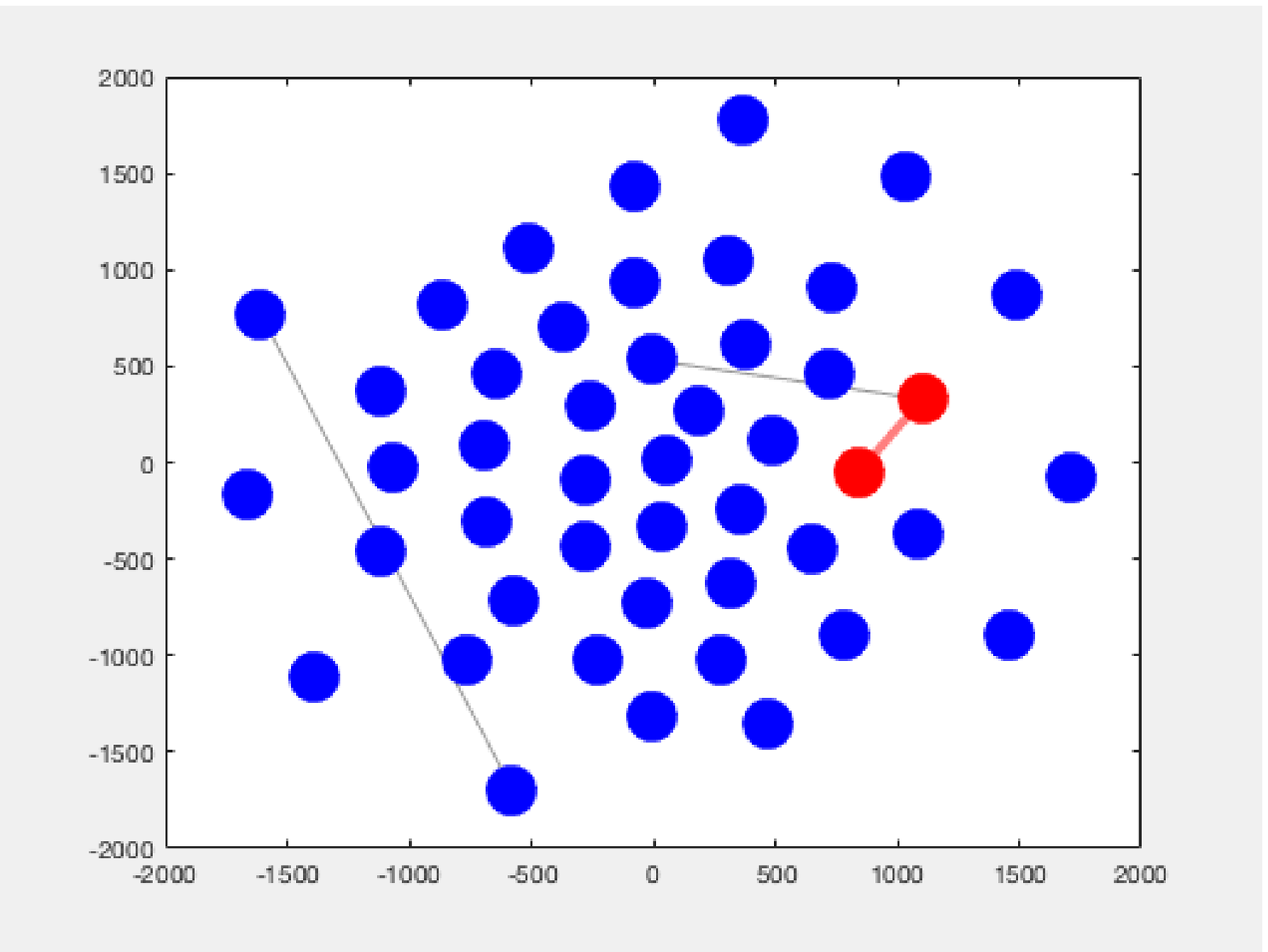}
\caption{Snapshot 18}
\end{subfigure}
\vspace{-0.2in}
\caption{2-D dynamic network vizualization of the Enron network. Nodes represent people in the Enron email network and edges represent an email between two people. We highlight the red nodes to show how our retrofitted model smoothly brings two nodes closer together before they form their first edge.}
\label{fig:dynamic-vis}
\vspace{-0.1in}
\end{figure*}

\subsection{Effect of Latent Dimensions} \label{sec:lat-dim}
In Figure \ref{fig:latentDimension}, we compare the performance of the baseline model (BCGD) with our best models for three different dataset (one representative dataset from each network group) and over three different latent dimension: 32, 64, and 128 using the NDCG metric. In all three datasets, the performance of multiple of our models is consistently better than BCGD over all three latent dimensions.
BCGD shows marginal improvement as the latent dimension increases. Most of our models stay flat or increase slowly as the dimension size increases because our models already achieve very high NDCG even in the low dimension (at dimension size, 32). The performance of RET (tsvd) and HeterLT (Node2Vec, avg) decease in DBLP2 and Facebook2, respectively as the dimension size increases from 64 to 128. However, the decrement is only around 2 points in both cases. To summarize, our models show
robust and consistently better performance than BCGD for a widely varying number
of latent dimensions.


\section{Dynamic Network Visualization}\label{sec:vis}
To demonstrate the smooth transition of the nodes between snapshots using the retrofitted model, we created an animation using the Enron dataset\footnotetext{https://www.cs.cmu.edu/~enron/}. The full video can be seen at: \url{https://www.youtube.com/watch?v=FtcaF0cv6iU}.
The dataset was divided into 18 equal-length time-stamps and the retrofitted model was applied. We then used TSNE to project the nodes into a 2-dimensional space so that they can be visualized as frames of an animation. Figure \ref{fig:dynamic-vis} shows the 2-dimensional network at time-stamps $t = 1, 11, 18$. The animation
demonstrates how the retrofitting model brings two faraway nodes (red
colored) in close proximity over time before an edge is created between them in the
final snapshot.

\section{Conclusion}
\label{sec:tempn2v:conclusion}
In this work we propose models for learning latent embedding vectors of 
vertices for all different temporal snapshots of a dynamic network. The
proposed models exploit temporal smoothing either at the node-level through
retrofitting, or at the network level through smooth linear transformation.
Extensive experiments over 9 dynamic networks from various domains show that
our proposed models generate superior vertex embedding than existing 
state-of-the-art methods for solving the task of temporal link prediction.
Visualization of embedding vectors over time shows the utility of the retrofitted
model for tracking the vertices over time to understand the evolution patterns of
a dynamic network.

\bibliographystyle{ACM-Reference-Format}
\bibliography{tempn2v} 

\end{document}